\documentclass[twoside,twocolumn]{article}

\def\titlename{Prospects for rare B decays at Belle~II}
\def\authorname{Sam Cunliffe}
\def\shortauthorname{S. Cunliffe}
\usepackage[english]{babel} 
\usepackage[hmarginratio=1:1,top=32mm,columnsep=20pt]{geometry} 
\usepackage[hang, small,labelfont=bf,up,textfont=it,up]{caption} 
\usepackage[nolist,nohyperlinks]{acronym} 
\usepackage{cite} 
\usepackage{graphicx}
\usepackage{multirow}
\usepackage{amsmath}
\usepackage{xspace}

\newcommand{\btwo}{{Belle~II}\xspace}
\newcommand{\epem}{\ensuremath{e^+e^-}\xspace}
\newcommand{\UpFS}{\ensuremath{\mathit{\Upsilon}(4S)}\xspace}
\newcommand{\UpFStoBBbar}{\ensuremath{\UpFS\to B\bar{B}}\xspace}
\newcommand{\invab}{ab$^{-1}$\xspace}
\newcommand{\gevsqcccc}{\ensuremath{\mathrm{GeV}^2/c^4}\xspace}

\newcommand{\ACP}{\ensuremath{A_{\rm CP}}\xspace}
\newcommand{\dACP}{\ensuremath{\Delta A_{\rm CP}}\xspace}
\newcommand{\AI}{\ensuremath{\Delta_{0+}}\xspace}
\newcommand{\eO}[1]{\ensuremath{\mathcal{O}_{#1}}\xspace}
\newcommand{\C}[1]{\ensuremath{C_{#1}}\xspace}
\newcommand{\CNP}[1]{\ensuremath{C_{#1}^{\textrm{NP}}}\xspace}
\newcommand{\eOi}{\eO{i}}
\newcommand{\Ci}{\C{i}}
\newcommand{\CiNP}{\CNP{i}}

\newcommand{\btosnn}{${b\to s\nu\bar{\nu}}$\xspace}
\newcommand{\btosll}{${b\to s\ell^+\ell^-}$\xspace}
\newcommand{\btosmm}{${b\to s\mu^+\mu^-}$\xspace}
\newcommand{\btoseemm}{${b\to s(e^+e^-,\,\mu^+\mu^-)}$\xspace}
\newcommand{\btosdtt}{${b\to (s,d)\tau^+\tau^-}$\xspace}

\newcommand{\btosdgamma}{${b\to (s,d)\gamma}$\xspace}

\newcommand{\BtoKstnn}{${B\to K^*\nu\bar{\nu}}$\xspace}
\newcommand{\BtoKstgamma}{${B\to K^*\gamma}$\xspace}
\newcommand{\Btott}{${B^0\to\tau^+\tau^-}$\xspace}
\newcommand{\Bstott}{${B_s\to\tau^+\tau^-}$\xspace}
\newcommand{\BtoXsdgamma}{${B\to X_{(s,\,d)}\gamma}$\xspace}
\newcommand{\BtoXsgamma}{${B\to X_s\gamma}$\xspace}
\newcommand{\BtoXdgamma}{${B\to X_d\gamma}$\xspace}

\newcommand{\BtoXsll}{${B\to X_s\ell^+\ell^-}$\xspace}
\newcommand{\BtoXsee}{${B\to Xe^+e^-}$\xspace}
\newcommand{\BtoXsmm}{${B\to X\mu^+\mu^-}$\xspace}
\newcommand{\BztoKstmm}{${B^0\to K^{*0}\mu^+\mu^-}$\xspace}
\newcommand{\BtoKstmm}{${B\to K^*\mu^+\mu^-}$\xspace}
\newcommand{\BtoKstee}{${B\to K^*e^+e^-}$\xspace}
\newcommand{\BtoKsteemm}{${B\to K^*(e^+e^-,\,\mu^+\mu^-)}$\xspace}
\newcommand{\BtoKtt}{${B^+\to K^+\tau^+\tau^-}$\xspace}

\usepackage{titlesec} 
\titleformat{\section}[block]{\normalsize\bf}{\thesection.}{0.5em}{}
\titleformat{\subsection}[block]{\normalsize\bf}{\thesubsection.}{0.5em}{}
\makeatletter
\g@addto@macro\bfseries{\boldmath}
\makeatother

\usepackage{fancyhdr}
\usepackage{color}
\definecolor{royalblue}{rgb}{0.25, 0.41, 0.88}
\usepackage{hyperref} 
\hypersetup{
  colorlinks,
  breaklinks,
  urlcolor=royalblue,
  linkcolor=royalblue,
  citecolor=royalblue
}

\usepackage{titling} 
\pretitle{\huge\bfseries}
\posttitle{\\}
\preauthor{\begin{flushleft}\large\bfseries}
\postauthor{\end{flushleft}\vspace*{-2\baselineskip}} 

\title{\titlename}
\author{\authorname\\
  \normalsize\it Pacific Northwest National Laboratory, 902 Battelle Blvd., Richland, WA 99352, USA. \\ 
  \rm\texttt{\href{mailto:samuel.cunliffe@pnnl.gov}{samuel.cunliffe@pnnl.gov}}\\
  orcid: {\href{https://orcid.org/0000-0003-0167-8641}{0000-0003-0167-8641}}\newline\\
  \normalsize Proceedings of the APS Division of Particles and Fields Meeting (DPF2017).\\
  \normalsize 29 July -- 4 August 2017, Fermilab, Batavia, IL, C170731.\newline\\
  Presented on behalf of the \btwo electroweak penguins working group.
  \newline\\
  \href{http://docs.belle2.org/record/669}{BELLE2-CONF-PROC-2017-028}; PNNL-SA-128762
}
\date{} 

\renewcommand{\maketitlehookd}{ 
  \begin{abstract}
    Rare and flavour-changing neutral current decays of the B meson are an important probe in the search for physics beyond the Standard Model. 
    There have recently been several anomalies in rare B decays, and lepton-universality measurements, specifically involving the \btosll quark transition. 
    These results tend towards a non-Standard-Model interpretation.
    The \btwo experiment is a next-generation $b$ physics experiment located at {SuperKEKB}, an upgraded B factory \epem collider, in Tsukuba, Japan. 
    The first collisions are expected in early 2018 with full physics data expected in 2019.
    This document describes prospects for several rare B decays at \btwo including \btosll processes and others, such as \btosdgamma and \btosnn. 
    Areas where the \btwo program is complementary to that of the currently running LHCb experiment are highlighted.
  \end{abstract}
}

\begin{document}

\maketitle

\section{Introduction}

\begin{figure}
  \centering
    \includegraphics[width=1.0\columnwidth]{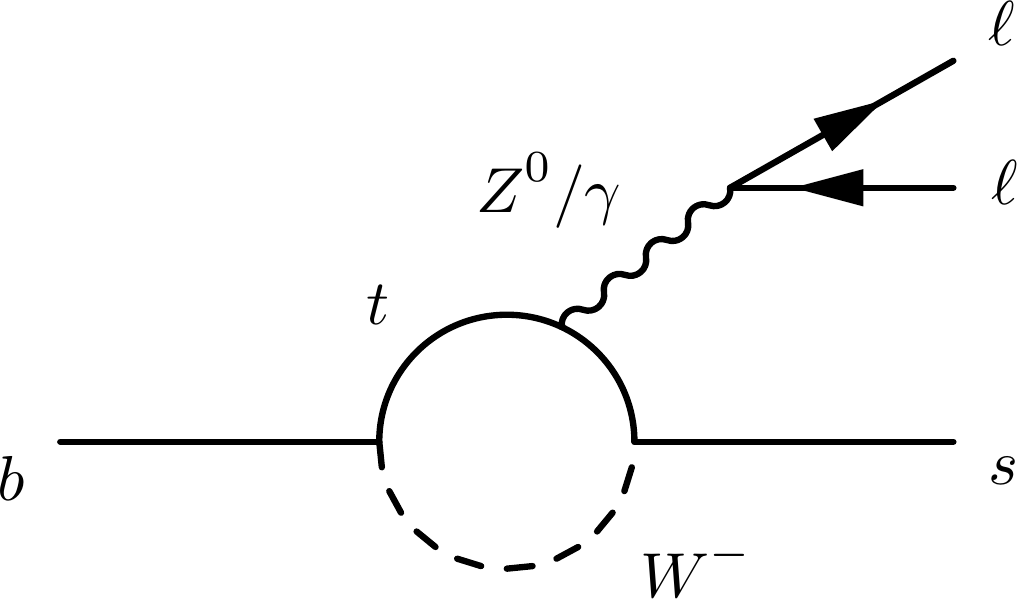}
    \includegraphics[width=1.0\columnwidth]{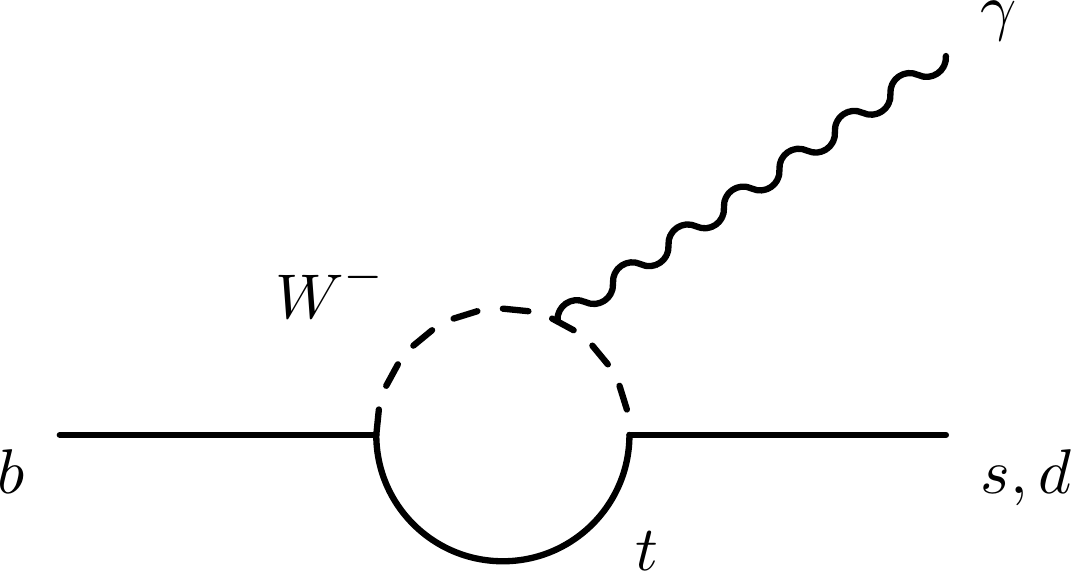}
    \caption{
      The leading order Feynman diagrams for the \ac{FCNC} $b\to s\ell\ell$ (where $\ell$ represents $e$, $\mu$, $\tau$ or $\nu$) and \btosdgamma in the \ac{SM}.
    \label{fig:btos_sm}
    }
\end{figure}

Rare and \ac{FCNC} decays of the beauty quark are sensitive to the effects of undiscovered new particles, if they exist, that are not included in the \ac{SM}.
As \ac{FCNC} are forbidden at tree-level in the \ac{SM}, these decays proceed by higher-order loop-level diagrams (as shown in Figure~\ref{fig:btos_sm}).
However any potential \ac{NP} contribution does not suffer the same restriction and can occur at a comparable size.
As these \ac{NP} diagrams contain virtual particles that can be off-mass-shell, the mass scale in the search for \ac{NP} with these decays is typically many times larger than searches involving direct production.
Furthermore, several recent measurements of these decay processes~\cite{LHCb-RK-2014,LHCb-RKst-2017,LHCb-Kstmm-2016,Belle-Kstmm-2017}\footnote{The result of Ref.~\cite{Belle-Kstmm-2017} is also presented at this conference, talk by S. Sandilya.} are in tension with \ac{SM} predictions, which has generated much interest~\cite{Glashow:2014iga,Altmannshofer:2017yso,Capdevila:2017bsm,Hiller:2017bzc,Sala:2017ihs}\footnote{The analysis of Ref.~\cite{Altmannshofer:2017yso} is also presented at this conference, talk by W. Altmannshofer.}.

The \btwo experiment~\cite{BelleIITDR} is a hermetic detector currently being commissioned at the {SuperKEKB} accelerator~\cite{SuperKEKBDesignPaper} at the KEK laboratory in Tsukuba, Japan.
The first data are expected in early 2018, with a target dataset of 50~\invab of \epem collisions by 2025.
This dataset should contain approximately $50\times10^{9}$ $e^+e^-\to\UpFStoBBbar$ events.
Measurements of rare and \ac{FCNC} decays are an integral part of the \btwo physics program.

\section{Previous and current $b$ physics experiments}

The \btwo experiment is a second-generation B factory.
The first generation of B factories were the Belle and BaBar experiments~\cite{PhysicsOfTheBFactories} at the KEKB and PEP-III accelerators at KEK and the SLAC National Accelerator Laboratory, respectively.
Following the first-generation B factories, the LHCb experiment~\cite{LHCb-detector-paper} at the LHC at CERN has been taking data since 2009.
The LHC produces $pp$ collisions at high-energy and cannot exploit the \UpFS resonance, however $pp\to b\bar b$ quark pair-production is copious in the forward region and LHCb utilises a single-arm forward spectrometer detector design~\cite{LHCb-detector-paper}.
These three experiments have produced many noteworthy results, several examples that are not discussed elsewhere in this document are described in Refs~\cite{BelleCPV,BaBarCPV,BelleX3872,BaBar-DstTauNu,CMSLHCb-Bsmm,LHCb-pentaquarks}.

\section{The next-generation experiment: \btwo  at {SuperKEKB}}

\subsection{Description}

The {SuperKEKB} accelerator complex~\cite{SuperKEKBDesignPaper} is an upgrade of KEKB.
The accelerator beams are asymmetric in energy: 7~GeV for electrons (defining the forward direction) and 4~GeV for positrons.  
The accelerator is designed to achieve a factor 40 increase in instantaneous luminosity with respect to KEKB.  
This is due to a more focused beam crossing and higher beam current, achieved by: new superconducting magnets~\cite{SuperKEKBFinalFocusMagnets} at the interaction point, a new positron dampening ring, and upgraded beam optics.

The \btwo detector~\cite{BelleIITDR} has been upgraded from Belle to cope with the much higher luminosity and higher expected beam backgrounds.
Around the collision point, the silicon vertex detector has one more layer than Belle, with the addition of two inner layers of depleted field-effect transistor pixel detectors.
The vertexing detectors are surrounded by the \btwo wire drift chamber that is larger than in Belle.
Two new particle identification systems have been installed utilising Cherenkov radiation: in aerogel blocks in the forward endcap and totally internally reflected inside quartz bars in the barrel.
The thallium-doped caesium iodide electromagnetic calorimeter has been reused from Belle although the readout and electronics have been totally replaced.
The superconducting coil magnet is reused and will provide a 1.5~T magnetic field for charge assignment and to measure tracking momentum of charged particles.
The inner barrel layers and the endcap of $K_L^0$ and muon detector have been replaced with plastic scintillator, the outer barrel layers reuse the original resistive plate chambers from Belle.

\subsection{Status as of August 2017}

\begin{figure*}
  \centering
    \includegraphics[width=0.7\textwidth]{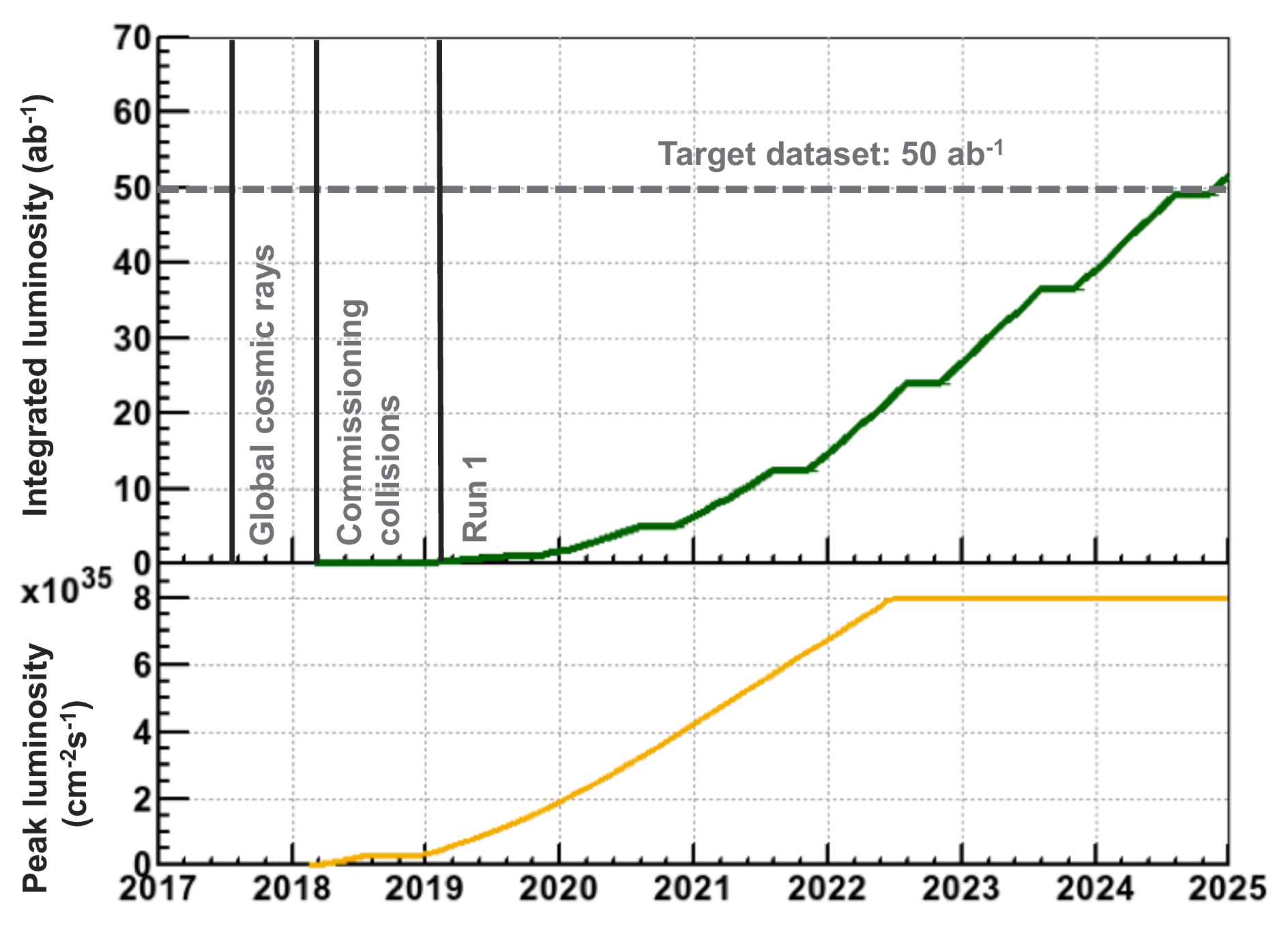}
    \caption{
      The scheduled integrated and peak luminosities of SuperKEKB.
      Approximately ${10^{9}}$ ${B\bar{B}}$ pairs per \invab of data at the \UpFS resonance will be collected.
      Adapted from~\cite{SuperKEKBWebsite}.
      \label{fig:SuperKEKBSchedule}
    }
\end{figure*}

The design data schedule is shown in Figure~\ref{fig:SuperKEKBSchedule}.
In early 2018 the accelerator is scheduled to provide \epem collisions at reduced luminosity for detector commissioning\footnote{Operationally, and in \btwo literature, this is referred to as ``Phase 2'' for historic reasons.}.
The vertexing pixel detector and silicon vertex detector components will not be installed for this phase.
In 2019 the full vertexing detector will be installed and the accelerator will provide physics collisions\footnote{Referred to as ``Phase 3''. Should be thought of as physics run 1.}.

\section{Theoretical framework and motivation}

Figure~\ref{fig:btos_sm} shows the leading order \ac{SM} contributions to $b\to s$ processes.
However, in order to interpret results in a model-independent way, the theory community typically works in a general expansion of an effective Hamiltonian,
\begin{equation*}
  \mathcal H_{\rm eff} \propto \sum_{i=1,..10,S,P}(\Ci\eOi+\Ci'\eOi'),
\end{equation*}
in terms of effective operators, $\eOi^{(\prime)}$, containing the non-pertubative low-energy effects and so-called ``Wilson coefficients'', \Ci.
The primes denote the chiral partner operator that is suppressed in the \ac{SM}.
The Wilson coefficients may be expressed as:
\begin{equation*}
  \Ci = \Ci^{\rm SM} + \CiNP,
\end{equation*}
the sum of the \ac{SM} (calculable with pertubative techniques) and \ac{NP} (to be determined) contributions.

\begin{figure}
  \centering
  \includegraphics[width=1.0\columnwidth]{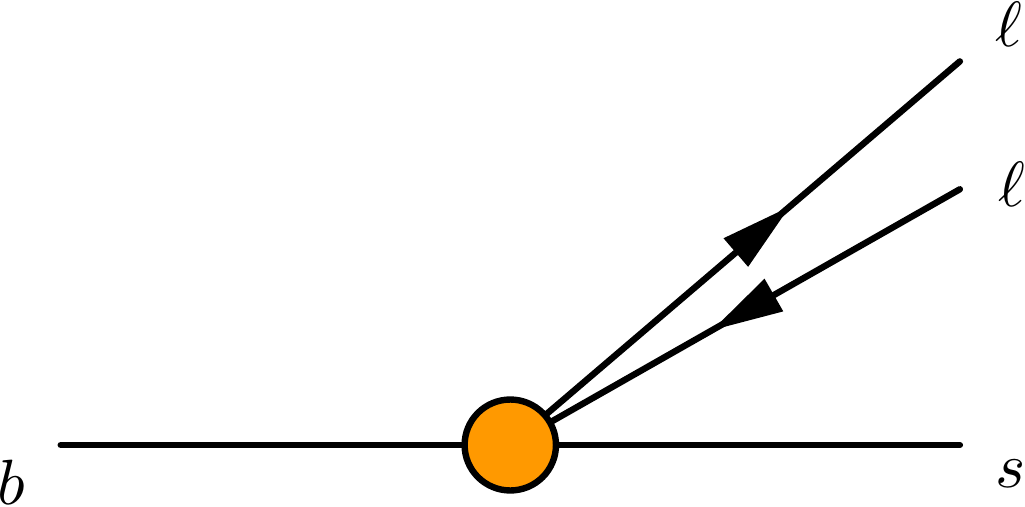}
  \newline\newline
  \includegraphics[width=1.0\columnwidth]{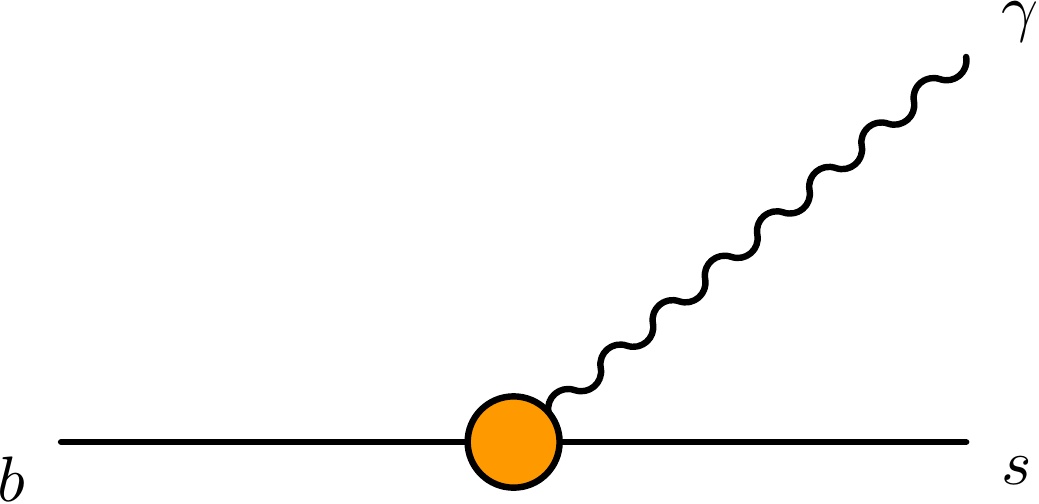}
  \caption{
    A diagrammatic representation of the effective (axial) vector operators \eO9 (\eO{10}) and radiative photon operator \eO7 (upper and lower respectively).
  \label{fig:operators}
  }
\end{figure}

There are 24 operators and coefficients in the full expansion\footnote{More detail is available in many references, for example Ref.~\cite{MannelEFTTextbook}.}. However, there are three which are most relevant for $b\to s$:
\begin{align*}
  \eO9  &\propto (\bar{s} \gamma_{\mu} P_L b)(\overline{\ell} \gamma^\mu \ell) ;\\
  \eO{10}  &\propto(\bar{s}  \gamma_{\mu} P_L b)(  \overline{\ell} \gamma^\mu \gamma_5 \ell) ;\\
  \eO7  &\propto m_b(\bar{s} \sigma_{\mu \nu} P_R b) F^{\mu \nu}.
\end{align*}
These are the vector, axial vector, and radiative photon operators respectively, and are shown schematically in Figure~\ref{fig:operators}.
This theoretical procedure is appealing as it allows model-independent global fits to the \CiNP, searching for {\it generic \ac{NP}} based on the {\it form of the interaction} with \ac{SM} fermions.

Furthermore, many observables are then constructed in such a way as to divide out or otherwise remove the hadronic uncertainties.
Such observables are called ``theoretically clean'' or ``form-factor independent''.
Finally, it is also possible to separate this whole framework by lepton flavour such that
\begin{equation*}
\Ci \longrightarrow \Ci^{\mu} \textrm{ and } \Ci^e
\end{equation*}
allowing for a lepton-flavour-dependent determination of any \ac{NP}.
This final step is relevant given recent results~\cite{LHCb-RK-2014,LHCb-RKst-2017,Belle-Kstmm-2017}, analyses may be found in Refs~\cite{Altmannshofer:2017yso,Capdevila:2017bsm,Hiller:2017bzc}.

\section{Inclusive measurements and full event interpretation}

Inclusive decays, where the hadronic part of the decay is not specified, are denoted ${B\to X(\gamma,\ell\ell)}$ with $X$ representing all hadrons.
Their measurements should be contrasted with measurements of exclusive decays to a specific hadronic final state (such as \BtoKstgamma, for example).

In terms of theory, \ac{SM} predictions of inclusive decays are complementary to exclusive as they typically suffer less, or orthogonal, uncertainty due to hadronic form-factors.
Predictions for the branching fractions of \BtoXsgamma and \BtoXsll in the \ac{SM} are calculated to a precision of around $7\%$~\cite{Huber:2007vv,Misiak:2015xwa,Paul:2016urs} which should be compared to the \ac{SM} prediction for the branching fractions of \BtoKstgamma and \BtoKstmm at $20-23\%$~\cite{BSZ,Paul:2016urs}.

Experimentally, there are two approaches for inclusive analyses: fully inclusive and the so-called ``sum-of-exclusive'' methods.
These are described in the following subsections.

In addition, decays with neutrinos in the final state are often reconstructed in the B factories as part of the full $\epem\to\UpFStoBBbar$ event such that the missing energy is precisely known.

\subsection{Sum-of-exclusive approach}

The sum-of-exclusive reconstruction method is where the $X$ is specifically reconstructed to several final states.
This method is the only way to specify the transition as $b\to s$ (or $b\to d$) since $X_s$ ($X_d$) can be specified by the presence (absence) of a kaon in the final state.
For example, $X_s$ can be reconstructed as $Kn\pi$ and $3Km\pi$ where $n\ge1$ and $m\ge0$.
Furthermore, the flavour and momentum of the parent $B$ is known without the need to perform full event interpretation (discussed in the following section).
The sum-of-exclusive analyses therefore have relatively high efficiency to select signal events.

\subsection{Full event interpretation}

\begin{figure}
  \centering
    \includegraphics[width=0.9\columnwidth]{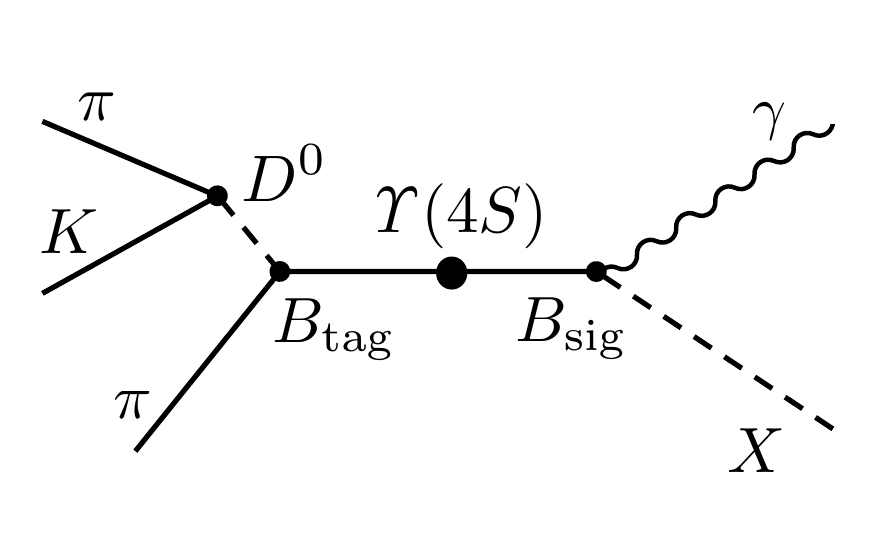}
    \caption{
      A schematic of a hadronically tagged decay where full event reconstruction is possible.
      In this example the tag decay is $B\to D^0(\to K\pi)\pi$, however several decay chains with high branching fractions or efficiencies are included~\cite{FEIKeck}.
      The signal is able to be constructed from missing energy (i.e. neutrinos) or as an inclusive decay if only a photon or (pair of) leptons are reconstructed.
      \label{fig:fei}
    }
\end{figure}

\begin{figure*}
  \centering
    \includegraphics[width=0.99\columnwidth]{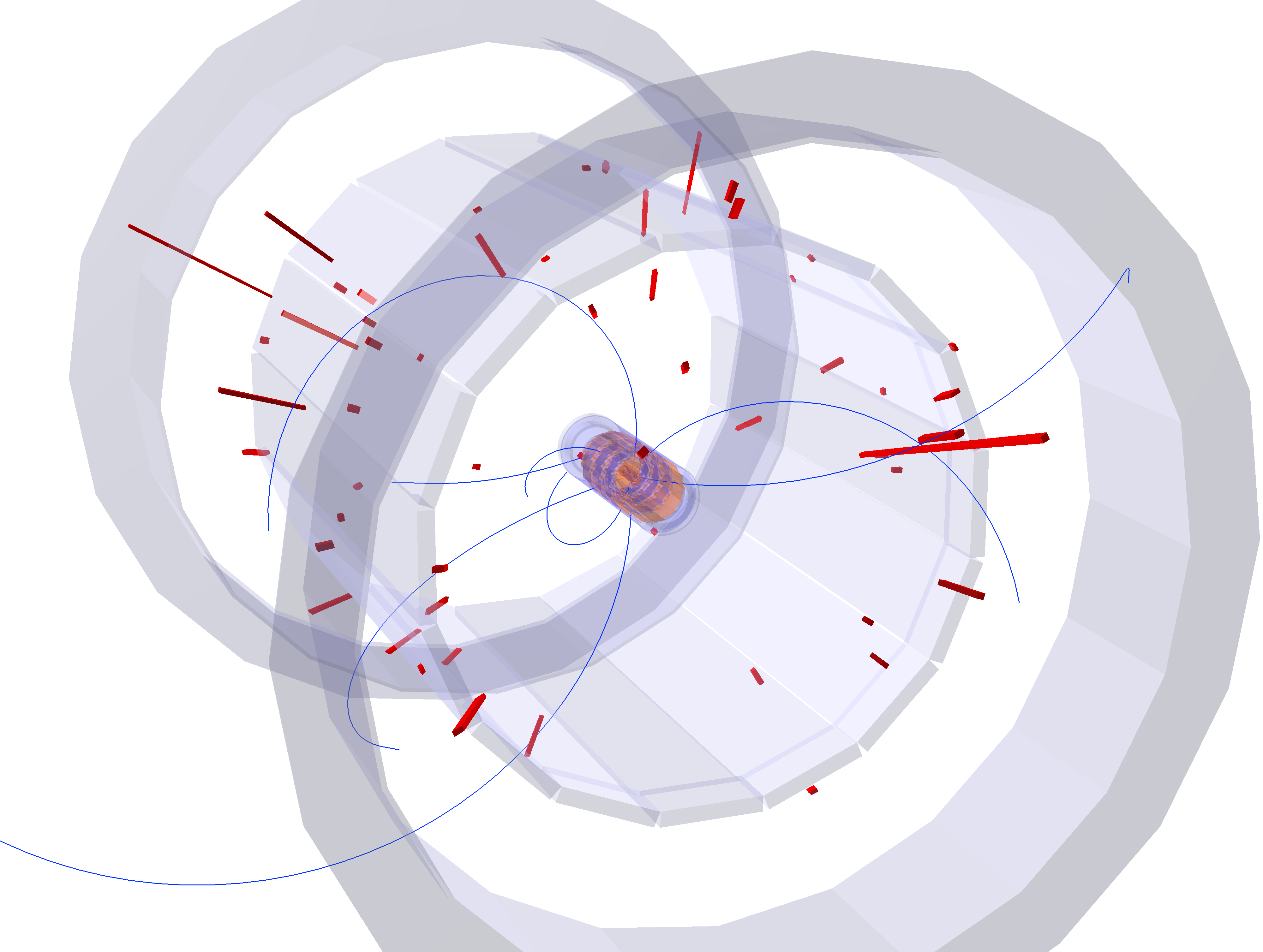}
    \includegraphics[width=0.99\columnwidth]{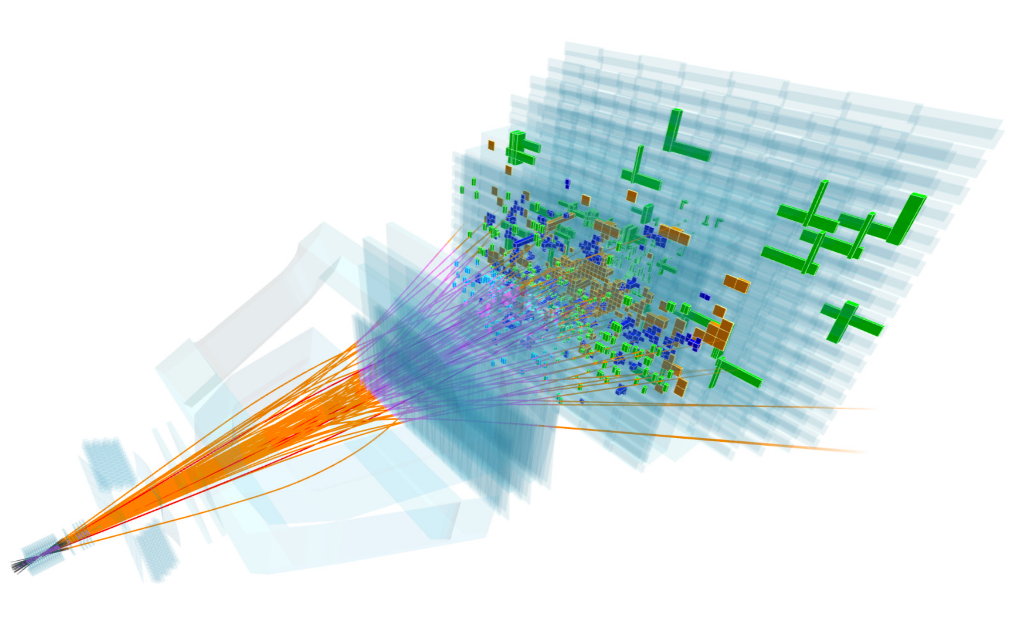}
    \caption{
      Examples of event displays from \btwo simulation (left) and LHCb data (right).
      \btwo is hermetic and can exploit the clean $\epem\to\UpFStoBBbar$ production.
      LHCb does not have full coverage and the production mechanism $p^+p^-\to b\bar{b}$ is less well known.
      \btwo is therefore better suited to perform fully inclusive measurements and measure processes with neutrinos in the final state.
      Right-hand figure from~\cite{LHCbEventDisplay}.
      \label{fig:detector-events}
    }
\end{figure*}

Fully inclusive measurements, and processes with neutrinos or other missing energy often rely on ``tagging'' the other $B$ decay (the $B_{\rm tag}$) from the \UpFS.
This is shown schematically in Figure~\ref{fig:fei}.
Such analyses are challenging for LHCb due to the detector geometry and the production mechanism $pp\to b\bar{b}$.
Hermetic detectors, such as BaBar, Belle, and \btwo are able to precisely reconstruct \UpFStoBBbar decays and therefore better suited to performing such measurements.
This is illustrated qualitatively in Figure~\ref{fig:detector-events}.

There are two further approaches to reconstruct the full event, dependent on the final state of the $B_{\rm tag}$ decay.
The $B_{\rm tag}$ decay may be reconstructed in a fully hadronic final state, in which case the momentum of the signal $B$ is known precisely, independent of the signal decay.
However, fully reconstructed hadronic decays suffer from low branching fractions with respect to semi-leptonic decays.
Thus semi-leptonic tagging has a somewhat higher efficiency at the expense of losing the full momentum (and parent charge) information in most signal decays.
Newer \ac{FEI} algorithms have been developed for \btwo~\cite{FEIKeck} which include several final states and utilise advances in machine learning.
The efficiencies of \ac{FEI} are shown in Table~\ref{tab:fei}.

\begin{table*}
  \begin{center}
    \begin{tabular}{lllll}
      \hline
      Parent                    & Tagging       & Belle II FEI & Belle MC Belle II FEI & Belle      \\ \hline
      \multirow{ 2}{*}{$B^\pm$} & Hadronic      & 0.61\%       & 0.49\%                & 0.28\%     \\
      ~                         & Semi-leptonic & 1.45\%       & 1.42\%                & 0.67\%     \\
      \multirow{2}{*}{$B^0$}    & Hadronic      & 0.34\%       & 0.33\%                & 0.18\%     \\
      ~                         & Semi-leptonic & 1.25\%       & 1.33\%                & 0.63\%     \\ \hline
    \end{tabular}
    \caption{
      Tagging efficiencies for \btwo \acs{FEI} algorithms determined with simulation of the \btwo and Belle detectors and the original Belle efficiency evaluated on data.
      Taken from~\cite{FEIKeck,b2tip}.
    }
    \label{tab:fei}
  \end{center}
\end{table*}

\section{Prospects at \btwo}

\subsection{Inclusive \texorpdfstring{\btosdgamma}{b-->(s,d)gamma}}

The inclusive radiative penguin measurements provide important constraints on many possible \ac{NP} scenarios such as models with extended Higgs sectors and supersymmetry~\cite{Paul:2016urs,Misiak:2017bgg}.
Measurements of the branching fraction of \BtoXsgamma (\BtoXdgamma) from the first generation of B factories have been combined~\cite{HFLAV} to give a precision of about 4\% (30\%).
These averages are in very good agreement with the  \ac{SM} predictions which are calculated to around 7\% precision~\cite{Misiak:2015xwa}.
With \btwo, the single experiment precision for \BtoXsgamma measured with a semi-leptonic tag is expected to quickly overtake the the combination, and should be known to less than 3\% uncertainty.
The full precision is expected to reach percent-level.
For \BtoXdgamma, the precision will be improved to around 15\% with 50~\invab.
In the case of $X_d$ this improvement is due both to a larger data sample, and to the upgraded particle identification system which will aid discrimination against the large background from \BtoXsgamma processes (namely where a kaon track is misidentified as a pion).

It is also possible to construct the CP-asymmetry, defined in general as:
\begin{equation*}
  \ACP \equiv 
  \frac{ \Gamma\left[\bar{B}\to\bar{f}\right] - \Gamma\left[B\to f\right] }
       { \Gamma\left[\bar{B}\to\bar{f}\right] + \Gamma\left[B\to f\right] },
\end{equation*}
where $\Gamma$ is the partial width for {\it any} decay of a $B$ meson to a final state $f$, as well as the isospin asymmetry, defined as:
\begin{equation*}
  \AI \equiv 
  \frac{ \Gamma\left[{B^0}\to{f^0}\right] - \Gamma\left[B^\pm\to f^\pm\right] }
    { \Gamma\left[{B^0}\to{f^0}\right] + \Gamma\left[B^\pm\to f^\pm\right] }.
\end{equation*}
Another related observable is the difference of CP-violation between the charged and neutral $B$ meson decays:
\begin{equation*}
  \dACP \equiv \ACP\left[B^\pm\to f^\pm\right] - \ACP\left[B^0\to f^0\right] .
\end{equation*}

These observables can be defined for an inclusive decay, such as \BtoXsgamma as,
\begin{align*}
  \ACP &=
  \frac{ \Gamma\left[\bar{B}\to X_s\gamma\right] - \Gamma\left[B\to X_{\bar{s}}\gamma\right] }
  { \Gamma\left[\bar{B}\to X_s\gamma\right] + \Gamma\left[B\to X_{\bar{s}}\gamma\right] }, \\
  \AI  &=
  \frac{ \Gamma\left[{B^0}\to{X_s}\gamma\right] - \Gamma\left[B^\pm\to X_s\gamma\right] }
       { \Gamma\left[{B^0}\to{X_s}\gamma\right] + \Gamma\left[B^\pm\to X_s\gamma\right] },\\
 \dACP &= \ACP\left[B^\pm\to X_s^\pm\gamma\right] - \ACP\left[B^0\to X_s^0\gamma\right].
\end{align*}
In all cases, the flavour and CP state of the parent $B$ is determined from the tag.

\begin{figure}
  \centering
    \includegraphics[width=1.0\columnwidth]{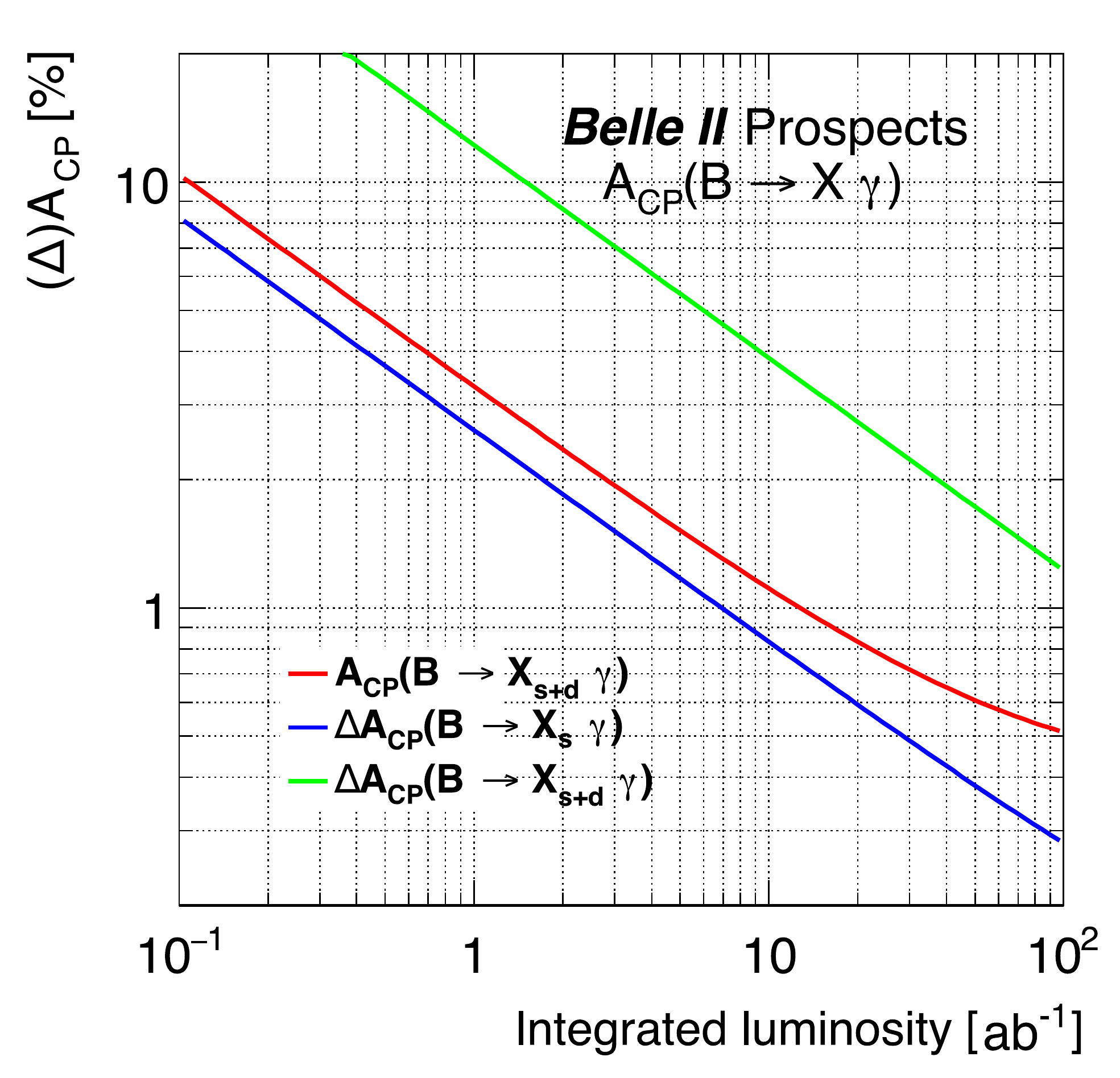}
    \caption{
    Sensitivity to \ACP and \dACP in \BtoXsdgamma decays.
    To appear in~\cite{b2tip}.
    \label{fig:ACP_BtoXGamma}
  }
\end{figure}

Such observables have reduced experimental systematic effects, as well as reduced theoretical uncertainty from hadronic form-factors.
Experimental measurements are therefore more precise than the branching fractions, for example \ACP and \AI for \BtoXsgamma are both around 2\%~\cite{Belle-BtoXgamma-2004,BaBar-BtoXgamma-2014}, for \BtoXdgamma they are around 30\%.
With 50~\invab at \btwo, measurements are expected to reach sub-percent-level precision for \ACP and \AI in \BtoXsgamma, and around percent level precision for \BtoXdgamma.
Figure~\ref{fig:ACP_BtoXGamma} shows the precision on \ACP and \dACP as a function of integrated luminosity collected at the \UpFS resonance.

\subsection{Lepton (non) universality and inclusive \texorpdfstring{\btoseemm}{b-->s(ee,mumu)}}

Recent experimental tests of lepton universality in \btosll decays have shown deviation from the \ac{SM} predictions~\cite{LHCb-RK-2014,LHCb-RKst-2017}.
Deviations are not too far from statistical significance and are therefore the source of much discussion within the community~\cite{Glashow:2014iga,Altmannshofer:2017yso,Capdevila:2017bsm,Hiller:2017bzc}.
In addition to these measurements a somewhat longstanding discrepancy in the angular analysis of \BztoKstmm~\cite{LHCb-Kstmm-2016} has been explored for both \BtoKstee and \BtoKstmm by Belle~\cite{Belle-Kstmm-2017}.

In global fits to the Wilson coefficients~\cite{Altmannshofer:2017yso,Capdevila:2017bsm,Hiller:2017bzc}, these discrepancies prefer a non-zero \CNP9.
In terms of \ac{NP} interpretations, models with an extended electroweak sector, such as a new vector boson $Z'$, have been suggested.
There has been some debate in the theory community about possible non-\ac{NP} explanations for these deviations, such as underestimated hadronic uncertainty, or an underestimated contribution from high-order diagrams involving charm quarks in the \btosmm transition~\cite{Lyon:2014hpa}.

\begin{figure}
  \centering
  \includegraphics[width=1.0\columnwidth]{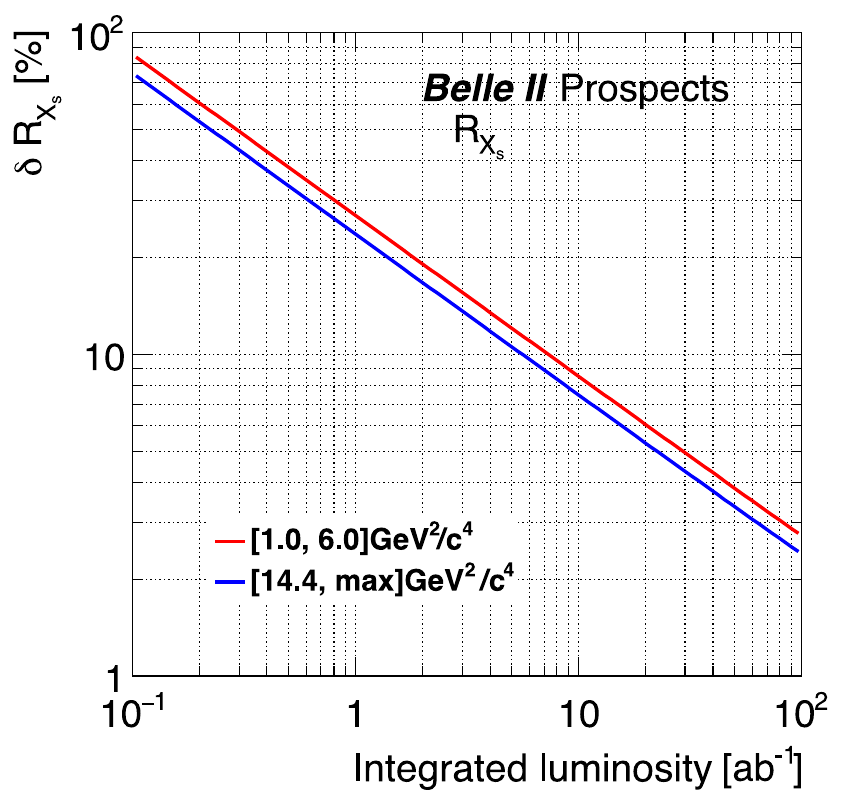}
  \caption{
    Sensitivity to an inclusive lepton universality ratio defined in Equation~\ref{eqn:RXs}, 
    for two regions of squared invariant mass of the lepton pair.
    To appear in~\cite{b2tip}.
    \label{fig:RXs}
  }
\end{figure}

References~\cite{LHCb-RK-2014,LHCb-RKst-2017} present the measurement of lepton universality ratios conventionally defined,
\begin{equation*}
  R_{K^{(*)}} \equiv \frac{\acs{BF}\left[B\to K^{(*)}\mu^+\mu^-\right]}
                          {\acs{BF}\left[B\to K^{(*)}e^+e^-\right]},
\end{equation*}
where \acs{BF} is the branching fraction.
In the \ac{SM} these ratios are predicted to be very close to unity within the region of the squared invariant mass of the lepton pair, ${1<q^2<6~\gevsqcccc}$~\cite{Hiller:2003js}.
\btwo will not overtake the precision of these measurements but will perform an independent verification.
With approximately 10~\invab (3~\invab) \btwo will reach the current precision of $R_{K}$ ($R_{K^*}$).
However an analogous definition in terms of the inclusive decays,
\begin{equation}\label{eqn:RXs}
  R_{X_s}\equiv \frac{\acs{BF}\left[B\to X_s\mu^+\mu^-\right]}
                          {\acs{BF}\left[B\to X_se^+e^-\right]},
\end{equation}
can be made.
Such an observable would be challenging for LHCb, but could be measured with percent-level precision at \btwo as shown in Figure~\ref{fig:RXs}.

\begin{figure}
  \centering
  \includegraphics[width=0.97\columnwidth]{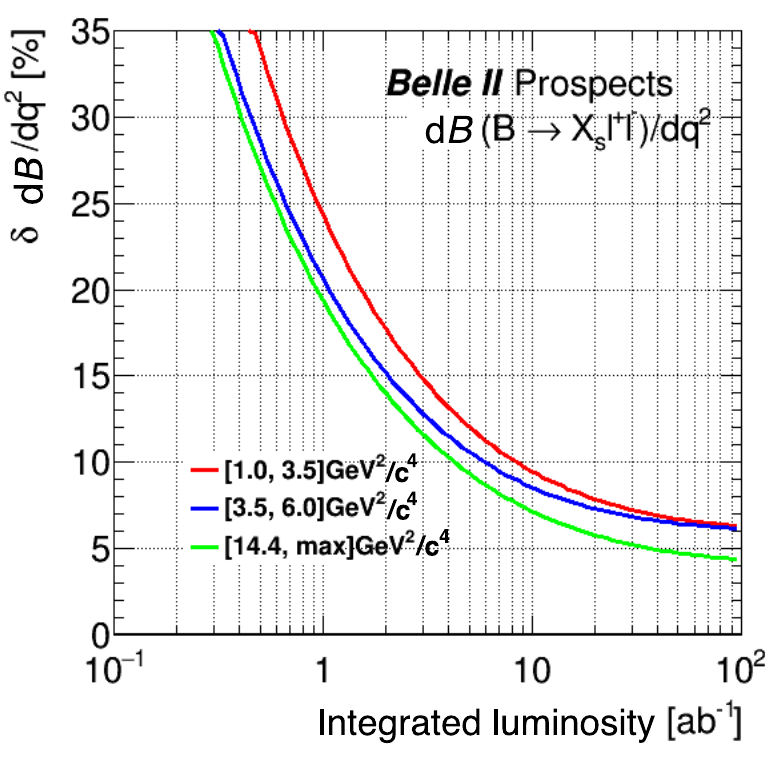}
  \caption{
    Sensitivity to the \acf{dBF} in \BtoXsll decays, for three regions of squared invariant mass of the lepton ($\ell= e,\,\mu$) pair.
    To appear in~\cite{b2tip}.
    \label{fig:BF_BtoXll}
  }
\end{figure}

\begin{figure}
  \centering
  \includegraphics[width=1.0\columnwidth]{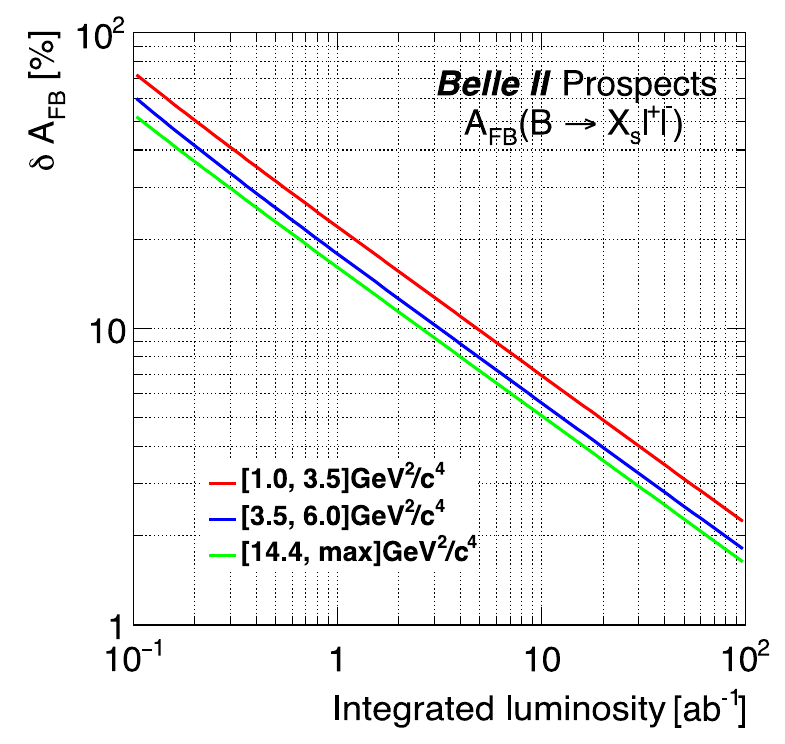}
  \caption{
    Sensitivity to \acf{AFB} in \BtoXsll decays, for three regions of squared invariant mass of the lepton ($\ell= e,\,\mu$) pair.
    To appear in~\cite{b2tip}.
    \label{fig:AFB_BtoXll}
  }
\end{figure}

It is also possible to measure the \ac{dBF}, \ACP, and perform an angular analysis for these inclusive \BtoXsee and \BtoXsmm decays.
In contrast to the angular analysis of the exclusive \BtoKstmm decay with many observables, in an inclusive angular analysis it is only possible to measure \ac{AFB}.
Current precision~\cite{Belle-BF-BtoXsll,BaBar-BFCPV-BtoXsll,Belle-AFB-BtoXsll} is around 30\% for \ac{dBF}, and 20\% for \ac{AFB} and \ACP.
\btwo will reach a precision of around 7\% for \ac{dBF} and $2-3\%$ for \ac{AFB} and \ACP.
Figures~\ref{fig:BF_BtoXll} and \ref{fig:AFB_BtoXll} show the sensitivity for the former two of these observables.

\subsection{\texorpdfstring{\btosnn}{b-->snunu}}

Assuming that the \BtoKstnn decay occurs at the rates predicted by the \ac{SM}~\cite{Buras:2014fpa,BELLE2-MEMO-2016-007},
\begin{align*}
  \acs{BF}\left[B^+\to K^{+}\nu\bar{\nu}\right] &= (4.7\pm0.6)\times10^{−6};  \\ 
  \acs{BF}\left[B^0\to K^{*0}\nu\bar{\nu}\right] &= (9.5\pm1.1)\times10^{−6},
\end{align*}
\btwo will observe the process and measure the branching fraction with $10-11\%$ uncertainty in 50~\invab.
This decay mode is of similar interest to \BztoKstmm in terms of sensitivity to \CNP{9,10}, however probing \BtoKstnn decays also provides orthogonal information.
For \BtoKstnn, the factorisation of hadronic effects is exact (since neutrinos are electrically neutral) and could be used to extract $B\to K$ hadronic form-factors to high accuracy~\cite{b2tip}.
It is also possible that \BtoKstnn can provide model-dependent information to disentangle possible \ac{NP} effects behind the current anomalies~\cite{Buras:2014fpa}.

\begin{figure}
  \centering
  \includegraphics[width=1.0\columnwidth]{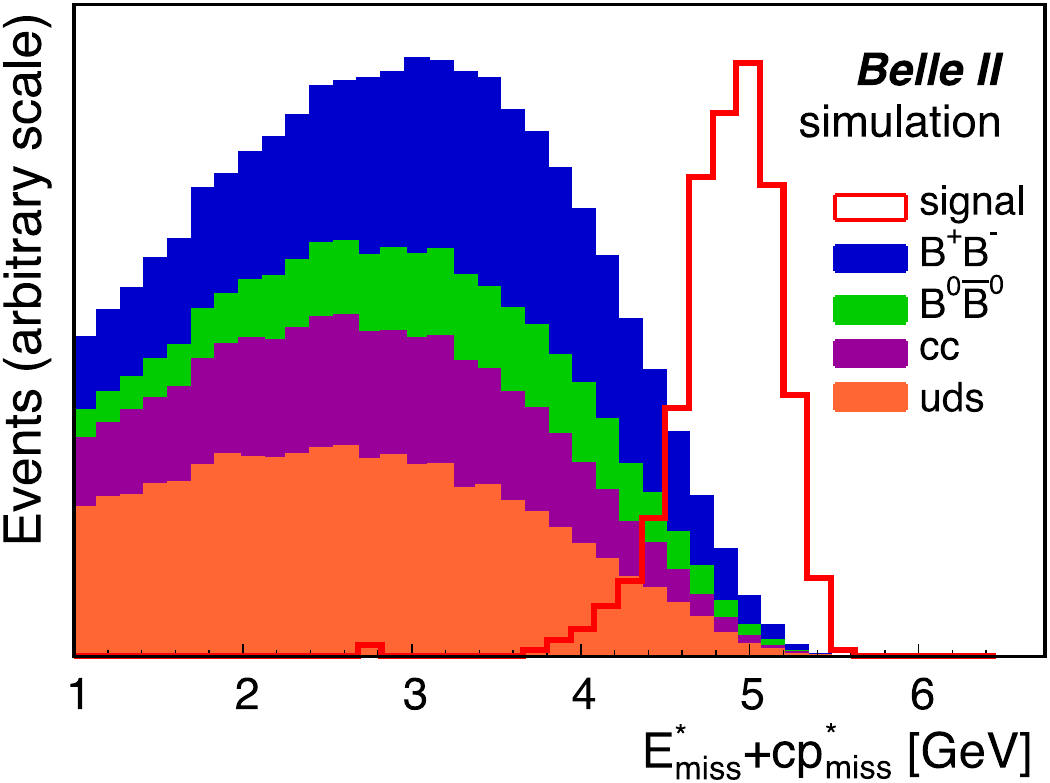}
  \caption{
    The distribution of the missing energy and missing momentum in the centre-of-momentum frame for \BtoKstnn signal decays (red) and various background categories (solid colour stack by cross-section).
    To appear in~\cite{b2tip}.
    \label{fig:BtoKstnn}
  }
\end{figure}

Experimentally, it is possible to use full event reconstruction and construct the sum of the missing energy and missing momentum in the \epem centre-of-momentum frame.
The distribution of this variable is shown in Figure~\ref{fig:BtoKstnn}.
Such a variable is promising for separating signal from background, either for a counting analysis or as the independent variable in a maximum likelihood fit.
Assuming observation at \btwo, it should also be possible to measure fraction of longitudinal polarisation of the $K^{*}$ in $B\to K^{*}\nu\bar\nu$ to around 20\% precision.

\subsection{\texorpdfstring{\btosdtt}{b-->(s,d)tautau}}

Decays with the \btosdtt transition are thus-far unobserved.
Current experimental limits~\cite{BaBar-BtoKtt,LHCb-Btott} are of the order of $10^{-3}$ which is rather far from the \ac{SM} predictions~\cite{Bobeth:2013uxa,Du:2015tda} of,
\begin{align*}
  \acs{BF}\left[B_s^0\to\tau^+\tau^-\right]   &= (7.73\pm0.49)\times10^{-7}; \\
  \acs{BF}\left[B^0\to\tau^+\tau^-\right]     &= (2.22\pm0.19)\times10^{-8}; \\
  \acs{BF}\left[B^+\to K^+\tau^+\tau^-\right] &= (1.22\pm0.10)\times10^{-7}.
\end{align*}
Assuming these \ac{SM} branching fractions, \btwo will be able to set limits of around $10^{-6}$ and $10^{-5}$ for \Btott and \BtoKtt respectively.
The sensitivity to \Bstott decays is highly dependent on SuperKEKB running at the $\mathit{\Upsilon}(5S)$ resonance, which has not yet been finalised.

\section{Conclusions}

The \btwo experiment at SuperKEKB will collect first collisions commissioning data in 2018.
Full-detector physics data are expected in 2019.
At time of writing, the \btwo detector has been rolled into the collision point at SuperKEKB and is taking cosmic ray commissioning data.

Rare radiative and electroweak penguin processes have recently shown deviations from \ac{SM} predictions, and form an integral part of the \btwo physics program.
\btwo will have access to several decay modes that are challenging at the LHCb experiment, such as \BtoKstnn and inclusive decays.
\btwo will provide independent verification of the deviations observed by LHCb, such as lepton universality in \BtoKsteemm decays.

\subsection*{Acknowledgements}

The author would like to thank Akimasa Ishikawa, Elisa Manoni, and Phillip Urquijo for working group collaboration and for their contributions to Ref.~\cite{b2tip} from which these proceedings draw upon.
The author is supported by US DOE grant FWP61210.

\begin{acronym}
  \acro{SM}[SM]{Standard Model of particle physics}
  \acro{NP}[NP]{new physics\acroextra{ (particles, and theories not included in the \acs{SM})}}
  \acro{FCNC}[FCNC]{flavour-changing neutral current}
  \acro{FEI}[FEI]{full event interpretation}
  \acro{BF}[\ensuremath{\mathcal{B}}\xspace]{branching fraction}
  \acro{dBF}[\ensuremath{\text{d}\mathcal{B}/\text{d}q^2}\xspace]{differential branching fraction}
  \acro{AFB}[\ensuremath{A_{\rm FB}}\xspace]{the forward-backward asymmetry of the leptons}
  \acro{AIunused}[\AI]{isospin asymmetry}
  \acro{ACPunused}[\ACP]{CP-asymmetry}
  \acro{dACPunused}[\dACP]{difference in CP-asymmetries between the charged and neutral modes}
\end{acronym}

\bibliography{main}

\providecommand{\href}[2]{#2}\begingroup\raggedright\begin{thebibliography}{10}

\bibitem{LHCb-RK-2014}
R.~Aaij et al., {\bf LHCb}, {\em {Test of lepton universality using
  $B^{+}\rightarrow K^{+}\ell^{+}\ell^{-}$ decays}\/},
  \href{http://dx.doi.org/10.1103/PhysRevLett.113.151601}{Phys. Rev. Lett. {\bf
  113} (2014)  151601},
\href{http://arxiv.org/abs/1406.6482}{{\tt arXiv:1406.6482}}.

\bibitem{LHCb-RKst-2017}
R.~Aaij et al., {\bf LHCb}, {\em {Test of lepton universality with $B^{0}
  \rightarrow K^{*0}\ell^{+}\ell^{-}$ decays}\/},
\href{http://arxiv.org/abs/1705.05802}{{\tt arXiv:1705.05802}}.

\bibitem{LHCb-Kstmm-2016}
R.~Aaij et al., {\bf LHCb}, {\em {Angular analysis of the $B^{0} \to K^{*0}
  \mu^{+} \mu^{-}$ decay using 3 fb$^{-1}$ of integrated luminosity}\/},
  \href{http://dx.doi.org/10.1007/JHEP02(2016)104}{JHEP {\bf 02} (2016)  104},
\href{http://arxiv.org/abs/1512.04442}{{\tt arXiv:1512.04442}}.

\bibitem{Belle-Kstmm-2017}
S.~Wehle et al., {\bf Belle}, {\em {Lepton-Flavor-Dependent Angular Analysis of
  $B\to K^\ast \ell^+\ell^-$}\/},
  \href{http://dx.doi.org/10.1103/PhysRevLett.118.111801}{Phys. Rev. Lett. {\bf
  118} (2017) 11, 111801},
\href{http://arxiv.org/abs/1612.05014}{{\tt arXiv:1612.05014}}.

\bibitem{Glashow:2014iga}
S.~L. Glashow et al., {\em {Lepton Flavor Violation in $B$ Decays?}\/},
  \href{http://dx.doi.org/10.1103/PhysRevLett.114.091801}{Phys. Rev. Lett. {\bf
  114} (2015)  091801},
\href{http://arxiv.org/abs/1411.0565}{{\tt arXiv:1411.0565}}.

\bibitem{Altmannshofer:2017yso}
W.~Altmannshofer et al., {\em {Interpreting Hints for Lepton Flavor
  Universality Violation}\/},
\href{http://arxiv.org/abs/1704.05435}{{\tt arXiv:1704.05435}}.

\bibitem{Capdevila:2017bsm}
B.~Capdevila et al., {\em {Patterns of New Physics in $b\to s\ell^+\ell^-$
  transitions in the light of recent data}\/},
\href{http://arxiv.org/abs/1704.05340}{{\tt arXiv:1704.05340}}.

\bibitem{Hiller:2017bzc}
G.~Hiller and I.~Nisandzic, {\em {$R_K$ and $R_{K^{\ast}}$ beyond the standard
  model}\/},  \href{http://dx.doi.org/10.1103/PhysRevD.96.035003}{Phys. Rev.
  {\bf D96} (2017) 3, 035003},
\href{http://arxiv.org/abs/1704.05444}{{\tt arXiv:1704.05444}}.

\bibitem{Sala:2017ihs}
F.~Sala and D.~M. Straub, {\em {A New Light Particle in B Decays?}\/},
\href{http://arxiv.org/abs/1704.06188}{{\tt arXiv:1704.06188}}.

\bibitem{BelleIITDR}
T.~Abe et al., {\bf Belle~II}, {\em {Belle~II Technical Design Report}\/},
\href{http://arxiv.org/abs/1011.0352}{{\tt arXiv:1011.0352}}.

\bibitem{SuperKEKBDesignPaper}
Y.~Ohnishi et al., {\em {Accelerator design at SuperKEKB}\/},
\href{http://dx.doi.org/10.1093/ptep/pts083}{Prog. Theo. Exp. Phys. {\bf 2013}
  (2013)  03A011}.

\bibitem{PhysicsOfTheBFactories}
A.~J. Bevan et al., {\bf Belle, BaBar}, {\em {The Physics of the B
  Factories}\/},  \href{http://dx.doi.org/10.1140/epjc/s10052-014-3026-9}{Eur.
  Phys. J. {\bf C74} (2014)  3026},
\href{http://arxiv.org/abs/1406.6311}{{\tt arXiv:1406.6311}}.

\bibitem{LHCb-detector-paper}
A.~A. Alves, Jr. et al., {\bf LHCb}, {\em {The LHCb Detector at the LHC}\/},
\href{http://dx.doi.org/10.1088/1748-0221/3/08/S08005}{JINST {\bf 3} (2008)
  S08005}.

\bibitem{BelleCPV}
K.~Abe et al., {\bf Belle}, {\em {Observation of large CP violation in the
  neutral $B$ meson system}\/},
  \href{http://dx.doi.org/10.1103/PhysRevLett.87.091802}{Phys. Rev. Lett. {\bf
  87} (2001)  091802},
\href{http://arxiv.org/abs/hep-ex/0107061}{{\tt arXiv:hep-ex/0107061}}.

\bibitem{BaBarCPV}
B.~Aubert et al., {\bf BaBar}, {\em {Observation of CP violation in the $B^0$
  meson system}\/},
  \href{http://dx.doi.org/10.1103/PhysRevLett.87.091801}{Phys. Rev. Lett. {\bf
  87} (2001)  091801},
\href{http://arxiv.org/abs/hep-ex/0107013}{{\tt arXiv:hep-ex/0107013}}.

\bibitem{BelleX3872}
S.~K. Choi et al., {\bf Belle}, {\em {Observation of a narrow charmonium-like
  state in exclusive $B^\pm \to K^\pm\pi^+\pi^- J/\psi$ decays}\/},
  \href{http://dx.doi.org/10.1103/PhysRevLett.91.262001}{Phys. Rev. Lett. {\bf
  91} (2003)  262001},
\href{http://arxiv.org/abs/hep-ex/0309032}{{\tt arXiv:hep-ex/0309032}}.

\bibitem{BaBar-DstTauNu}
J.~P. Lees et al., {\bf BaBar}, {\em {Evidence for an excess of $\bar{B} \to
  D^{(*)} \tau^-\bar{\nu}_\tau$ decays}\/},
  \href{http://dx.doi.org/10.1103/PhysRevLett.109.101802}{Phys. Rev. Lett. {\bf
  109} (2012)  101802},
\href{http://arxiv.org/abs/1205.5442}{{\tt arXiv:1205.5442}}.

\bibitem{CMSLHCb-Bsmm}
V.~Khachatryan et al., {\bf LHCb, CMS}, {\em {Observation of the rare
  $B^0_s\to\mu^+\mu^-$ decay from the combined analysis of CMS and LHCb
  data}\/},  \href{http://dx.doi.org/10.1038/nature14474}{Nature {\bf 522}
  (2015)  68--72},
\href{http://arxiv.org/abs/1411.4413}{{\tt arXiv:1411.4413}}.

\bibitem{LHCb-pentaquarks}
R.~Aaij et al., {\bf LHCb}, {\em {Observation of $J/\psi p$ Resonances
  Consistent with Pentaquark States in $\mathit{\Lambda}_b^0 \to J/\psi K^- p$
  Decays}\/},  \href{http://dx.doi.org/10.1103/PhysRevLett.115.072001}{Phys.
  Rev. Lett. {\bf 115} (2015)  072001},
\href{http://arxiv.org/abs/1507.03414}{{\tt arXiv:1507.03414}}.

\bibitem{SuperKEKBFinalFocusMagnets}
N.~Ohuchi et al., {\em {Design and Construction of the SuperKEKB QC1 Final
  Focus Superconducting Magnets}\/},
\href{http://dx.doi.org/10.1109/TASC.2014.2364858}{IEEE Trans. Appl. Supercond.
  {\bf 25} (2015) 3, 4001204}.

\bibitem{SuperKEKBWebsite}
\url{http://www-superkekb.kek.jp/}.

\bibitem{MannelEFTTextbook}
T.~Mannel, {\em {Effective Field Theories in Flavor Physics}\/},
\href{http://dx.doi.org/10.1007/b62268}{Springer Tracts Mod. Phys. {\bf 203}
  (2004)  }.

\bibitem{Huber:2007vv}
T.~Huber et al., {\em {Logarithmically Enhanced Corrections to the Decay Rate
  and Forward Backward Asymmetry in $\bar{B} \to X_s \ell^+ \ell^-$}\/},
  \href{http://dx.doi.org/10.1016/j.nuclphysb.2008.04.028}{Nucl. Phys. {\bf
  B802} (2008)  40--62},
\href{http://arxiv.org/abs/0712.3009}{{\tt arXiv:0712.3009}}.

\bibitem{Misiak:2015xwa}
M.~Misiak et al., {\em {Updated NNLO QCD predictions for the weak radiative
  B-meson decays}\/},
  \href{http://dx.doi.org/10.1103/PhysRevLett.114.221801}{Phys. Rev. Lett. {\bf
  114} (2015) 22, 221801},
\href{http://arxiv.org/abs/1503.01789}{{\tt arXiv:1503.01789}}.

\bibitem{Paul:2016urs}
A.~Paul and D.~M. Straub, {\em {Constraints on new physics from radiative $B$
  decays}\/},  \href{http://dx.doi.org/10.1007/JHEP04(2017)027}{JHEP {\bf 04}
  (2017)  027},
\href{http://arxiv.org/abs/1608.02556}{{\tt arXiv:1608.02556}}.

\bibitem{BSZ}
A.~Bharucha et al., {\em {$B\to V\ell^+\ell^-$ in the Standard Model from
  light-cone sum rules}\/},
  \href{http://dx.doi.org/10.1007/JHEP08(2016)098}{JHEP {\bf 08} (2016)  098},
\href{http://arxiv.org/abs/1503.05534}{{\tt arXiv:1503.05534}}.

\bibitem{FEIKeck}
T.~Keck, {\em The Full Event Interpretation for Belle II\/},  masters thesis,
  Karlsruher Institut f{\"u}r Technologie, 2014.
\newblock
  \href{https://ekp-invenio.physik.uni-karlsruhe.de/record/48602}{EKP-2015-00001}.

\bibitem{LHCbEventDisplay}
\url{https://lbevent.cern.ch/EventDisplay/index.html}.

\bibitem{b2tip}
E.~Kou, P.~Urquijo, et al., {\bf Belle~II}, {\em {The Belle~II Physics
  Book}\/},  {Under preparation, to be submitted to Prog. Theor. Exp. Phys.
  (2017)}.

\bibitem{Misiak:2017bgg}
M.~Misiak and M.~Steinhauser, {\em {Weak radiative decays of the B meson and
  bounds on $M_{H^\pm }$ in the Two-Higgs-Doublet Model}\/},
  \href{http://dx.doi.org/10.1140/epjc/s10052-017-4776-y}{Eur. Phys. J. {\bf
  C77} (2017) 3, 201},
\href{http://arxiv.org/abs/1702.04571}{{\tt arXiv:1702.04571}}.

\bibitem{HFLAV}
Y.~Amhis et al., {\bf HFLAV}, {\em {Averages of $b$-hadron, $c$-hadron, and
  $\tau$-lepton properties as of summer 2016}\/},
  \href{http://arxiv.org/abs/1612.07233}{{\tt arXiv:1612.07233}}.
Latest values available online at
  \url{http://www.slac.stanford.edu/xorg/hflav}.

\bibitem{Belle-BtoXgamma-2004}
S.~Nishida et al., {\bf Belle}, {\em {Measurement of the CP asymmetry in $B \to
  X_s \gamma$}\/},
  \href{http://dx.doi.org/10.1103/PhysRevLett.93.031803}{Phys. Rev. Lett. {\bf
  93} (2004)  031803},
\href{http://arxiv.org/abs/hep-ex/0308038}{{\tt arXiv:hep-ex/0308038}}.

\bibitem{BaBar-BtoXgamma-2014}
J.~P. Lees et al., {\bf BaBar}, {\em {Measurements of direct CP asymmetries in
  $B\to X_s\gamma$ decays using sum of exclusive decays}\/},
  \href{http://dx.doi.org/10.1103/PhysRevD.90.092001}{Phys. Rev. {\bf D90}
  (2014) 9, 092001},
\href{http://arxiv.org/abs/1406.0534}{{\tt arXiv:1406.0534}}.

\bibitem{Lyon:2014hpa}
J.~Lyon and R.~Zwicky, {\em {Resonances gone topsy turvy - the charm of QCD or
  new physics in $b \to s \ell^+ \ell^-$?}\/},
\href{http://arxiv.org/abs/1406.0566}{{\tt arXiv:1406.0566}}.

\bibitem{Hiller:2003js}
G.~Hiller and F.~Kruger, {\em {More model-independent analysis of $b \to s$
  processes}\/},  \href{http://dx.doi.org/10.1103/PhysRevD.69.074020}{Phys.
  Rev. {\bf D69} (2004)  074020},
\href{http://arxiv.org/abs/hep-ph/0310219}{{\tt arXiv:hep-ph/0310219}}.

\bibitem{Belle-BF-BtoXsll}
M.~Iwasaki et al., {\bf Belle}, {\em {Improved measurement of the electroweak
  penguin process $B \to X_s \ell^+ \ell^-$}\/},
  \href{http://dx.doi.org/10.1103/PhysRevD.72.092005}{Phys. Rev. {\bf D72}
  (2005)  092005},
\href{http://arxiv.org/abs/hep-ex/0503044}{{\tt arXiv:hep-ex/0503044}}.

\bibitem{BaBar-BFCPV-BtoXsll}
J.~P. Lees et al., {\bf BaBar}, {\em {Measurement of the $B \to X_s
  \ell^+\ell^-$ branching fraction and search for direct CP violation from a
  sum of exclusive final states}\/},
  \href{http://dx.doi.org/10.1103/PhysRevLett.112.211802}{Phys. Rev. Lett. {\bf
  112} (2014)  211802},
\href{http://arxiv.org/abs/1312.5364}{{\tt arXiv:1312.5364}}.

\bibitem{Belle-AFB-BtoXsll}
Y.~Sato et al., {\bf Belle}, {\em {Measurement of the lepton forward-backward
  asymmetry in $B \rightarrow X_s \ell^+ \ell^-$ decays with a sum of exclusive
  modes}\/},  \href{http://dx.doi.org/10.1103/PhysRevD.93.032008}{Phys. Rev.
  {\bf D93} (2016) 3, 032008}, \href{http://arxiv.org/abs/1402.7134}{{\tt
  arXiv:1402.7134}}.
{\href{https://doi.org/10.1103/PhysRevD.93.059901}{Phys. Rev. {\bf D93} (2016),
  5, 059901}}.

\bibitem{Buras:2014fpa}
A.~J. Buras et al., {\em {$ B\to {K}^{\left(\ast \right)}\nu \bar{\nu} $ decays
  in the Standard Model and beyond}\/},
  \href{http://dx.doi.org/10.1007/JHEP02(2015)184}{JHEP {\bf 02} (2015)  184},
\href{http://arxiv.org/abs/1409.4557}{{\tt arXiv:1409.4557}}.

\bibitem{BELLE2-MEMO-2016-007}
D.~M. Straub, {\em {$B\to {K}^{\left(\ast \right)}\nu \bar{\nu} $ SM
  predictions}\/},  2015,
  \href{https://docs.belle2.org/record/328}{BELLE2-MEMO-2016-007}.

\bibitem{BaBar-BtoKtt}
J.~P. Lees et al., {\bf BaBar}, {\em {Search for $B^{+}\rightarrow K^{+}
  \tau^{+}\tau^{-}$ at the BaBar experiment}\/},
  \href{http://dx.doi.org/10.1103/PhysRevLett.118.031802}{Phys. Rev. Lett. {\bf
  118} (2017) 3, 031802},
\href{http://arxiv.org/abs/1605.09637}{{\tt arXiv:1605.09637}}.

\bibitem{LHCb-Btott}
K.~De~Bruyn, {\bf LHCb}, {\em {Search for the rare decays
  $B^0_{(s)}\to\tau^+\tau^-$}\/},  2016,
  \href{http://cds.cern.ch/record/2220757}{LHCb-CONF-2016-011}.

\bibitem{Bobeth:2013uxa}
C.~Bobeth et al., {\em {$B_{s,d} \to \ell^+ \ell^-$ in the Standard Model with
  Reduced Theoretical Uncertainty}\/},
  \href{http://dx.doi.org/10.1103/PhysRevLett.112.101801}{Phys. Rev. Lett. {\bf
  112} (2014)  101801},
\href{http://arxiv.org/abs/1311.0903}{{\tt arXiv:1311.0903}}.

\bibitem{Du:2015tda}
D.~Du et al., {\em {Phenomenology of semileptonic B-meson decays with form
  factors from lattice QCD}\/},
  \href{http://dx.doi.org/10.1103/PhysRevD.93.034005}{Phys. Rev. {\bf D93}
  (2016) 3, 034005},
\href{http://arxiv.org/abs/1510.02349}{{\tt arXiv:1510.02349}}.

\end{thebibliography}\endgroup
\bibliographystyle{belle2-note}

\end{document}